# The Rise of Generative AI for Metal–Organic Framework Design and Synthesis

Chenru Duan,[1,2,*] Aditya Nandy,[1,3,*] Shyam Chand Pal,[4,5] Xin Yang,[6] Wenhao Gao,[7] Yuanqi Du,[8] Hendrik Kraß,[9] Yeonghun Kang,[10,11] Varinia Bernales,[10,12] Zuyang Ye,[13] Tristan Pyle,[14] Ray Yang,[4,5] Zeqi Gu,[8] Philippe Schwaller,[15] Shengqian Ma,[14] Shijing Sun,[13] Alán Aspuru-Guzik,[9,10,11,12,16,17,18,19] Seyed Mohamad Moosavi,[9] Robert Wexler,[4,5] Zhiling Zheng[1,4,5,*]

[1]These authors contributed equally to this work

[2]Deep Principle, Inc., Cambridge, MA, 02139, United States

[3]Department of Chemical and Biomolecular Engineering, University of California, Los Angeles, Los Angeles, CA 90095, United States

[4]Department of Chemistry, Washington University, St. Louis, MO 63130, United States

[5]Institute of Materials Science & Engineering, Washington University, St. Louis, Missouri 63130, United States

[6]David H. Koch Institute for Integrative Cancer Research, Massachusetts Institute of Technology, Cambridge, MA 02139, United States

[7]Department of Chemical Engineering, Massachusetts Institute of Technology, Cambridge, MA 02139, United States

[8]Department of Computer Science, Cornell University, Ithaca, NY 14853, United States

[9]Department of Chemical Engineering and Applied Chemistry, University of Toronto, Toronto, ON M5S 3E5, Canada

[10]Department of Chemistry, University of Toronto, Toronto, ON, M5S 3H6, Canada

[11]Vector Institute for Artificial Intelligence, Toronto, ON M5G 0C6, Canada

[12]Acceleration Consortium, University of Toronto, Toronto, ON M5G 1X6, Canada

[13]Department of Mechanical Engineering, University of Washington, Seattle, WA 98195, United States

[14]Department of Chemistry, University of North Texas, Denton, Texas 76201, United States

[15]Laboratory of Artificial Chemical Intelligence, École Polytechnique Fédérale de Lausanne, Lausanne 1015, Switzerland

[16]Department of Computer Science, University of Toronto, Toronto, ON M5S 1A1, Canada

[17]Department of Materials Science & Engineering, University of Toronto, 184 College St., Toronto, ON M5S 3E4, Canada

[18]Senior Fellow, Canadian Institute for Advanced Research, 661 University Ave., Toronto, ON M5G 1M1, Canada

[19]NVIDIA, 431 King St W #6th, Toronto, ON M5V 1K4, Canada

*E-mail: z.z@wustl.edu; duanchenru@gmail.com; aditya.nandy@ucla.edu

**Abstract**

Advances in generative artificial intelligence are transforming how metal–organic frameworks (MOFs) are designed and discovered. This Perspective introduces the shift from laborious enumeration of MOF candidates to generative approaches that can autonomously propose and synthesize in the laboratory new porous reticular structures on demand. We outline the progress of employing deep learning models, such as variational autoencoders, diffusion models, and large language model-based agents, that are fueled by the growing amount of available data from the MOF community and suggest novel crystalline materials designs. These generative tools can be combined with high-throughput computational screening and even automated experiments to form accelerated, closed-loop discovery pipelines. The result is a new paradigm for reticular chemistry in which AI algorithms more efficiently direct the search for high-performance MOF materials for clean air and energy applications. Finally, we highlight remaining challenges such as synthetic feasibility, dataset diversity, and the need for further integration of domain knowledge.

**1. Introduction**

Reticular synthesis of metal–organic frameworks (MOFs) epitomizes a materials design paradigm where modular inorganic nodes and organic linkers are linked into crystalline networks to engineer a periodic structure at the atomic level.[1,2] Since the concept of reticular chemistry was proposed after the synthesis of the first MOFs[3,4], the field has exploded with more than 100,000 distinct MOF structures synthesized to date,[5,6] leading to a wide range of programmable materials with novel properties. Notably, the global MOF market continues to expand, with commercial demand projected to reach several hundred tonnes annually, and multiple companies have already deployed MOFs at industrial scale as solid sorbents.[7,8] However, This immense structural diversity is both a triumph and a challenge: researchers can, in principle, create an almost boundless number of MOFs by swapping building blocks and topologies, yet identifying the optimal material for a given application is very similar to finding a needle in an infinitely large haystack due to the intractable combinatorics of MOF chemical space. Early efforts tackled this search by systematic enumeration.[9,10] While such template-based approaches provided invaluable databases for gas storage and separation studies, they inevitably explored only a tiny fraction of the vast "MOF universe." Indeed, it is estimated that millions of MOF structures have been predicted *in silico* beyond those made experimentally.[11] Societal needs like carbon capture,[12,13] clean water,[6,14] clean energy,[15–17] and catalysis[18,19] may be met by reticular materials with exceptional performance that lie in uncharted regions of chemical space.[2,20] Thus, the following question arises: *how can we intelligently navigate this exponentially large design space to discover MOFs with transformative properties for desired applications*?

The recent revolution in deep generative artificial intelligence (GenAI) models, ranging from variational autoencoders (VAEs)[21] to generative adversarial networks (GANs)[22], diffusion models[21] and language models[24], offers an unprecedented opportunity to "dream up" novel reticular materials *in silico* and subsequently realize them in the laboratory.[2,25–30] Broadly speaking, GenAI models are machine learning models that learn the underlying patterns of data to create new, similar data.[31] While traditional trial-and-error approaches struggle to keep up with the sheer combinatorial possibilities offered by reticular chemistry, generative models offer a way to unlock creativity at the atomic scale, proposing candidates that meet multiple design objectives simultaneously (e.g. high gas uptake and selectivity/stability / reduced cost).[2,11,28,29,32–36] In this Perspective, we explore how generative AI is reshaping MOF discovery, melding computational prediction with experimental realization (**Figure 1**). We first discuss the evolution of generative techniques for crystalline materials and how they were adapted to MOFs. We then highlight specific strategies, which range from fragment-based assembly to 3D diffusion, that address the unique challenges of MOF generation. Next, we examine computational workflows and platforms that integrate generative models with high-throughput screening and even autonomous labs, closing the loop between virtual and real-world discovery. A key focus is to demonstrate how machine imagination can propose MOFs beyond human intuition, how these proposals are vetted through physics-based simulations and experiments, and how a new generation of AI tools and foundation models is accelerating the entire workflow of materials discovery.

## 2. From Enumerative Design to Generative Modeling

In the early 2010s, the dominant computer-assisted approach to discovering new MOFs was enumerative design: researchers built hypothetical MOFs by stitching together known molecular components in all allowable ways. For example, an early study by Wilmer et al. assembled MOFs from libraries of secondary building units (SBUs) and linkers according to known nets, producing a "hypothetical MOF" database that could be screened for high methane storage capacity.[9] Such databases yielded valuable insights as top candidates from screening pre-existing building blocks and network architectures were later synthesized and confirmed as adsorbents (**Figure 1**).[37] Efforts were subsequently taken by Colón et al. to diversify the enumerated topologies.[38] Such efforts on topology screening have demonstrated success in identifying MOFs with high $CO_2$ adsorption capacity and concomitant selectivity in a flue gas stream.[39] The next leap was to accelerate the diversity and efficiency of building blocks assembly and potentially remove those topological limitations in order to let an algorithm imagine entirely new MOFs not confined to prior templates. This required a fundamentally different approach: generative models that learn the underlying rules of assembly from data and then extrapolate to create novel structures.

Generative modeling of crystalline materials is a non-trivial problem. Unlike molecules (which can be described by finite graphs) or images (fixed-size pixel grids), a MOF is an infinite periodic structure with long-range order and hundreds to thousands of atoms in a unit cell (**Figure 1b**). Successes in generative AI for materials occurred in similar systems such as zeolites, a class of inorganic porous crystals with analogies to MOFs. In 2020, Kim *et al.* reported an approach to generate zeolite structures using a generative adversarial network, dubbed "ZeoGAN".[40] They represented zeolites on a 3D grid to capture the pore structure of the crystal and trained a GAN to produce new frameworks with targeted methane adsorption energies. This was one of the first demonstrations that a neural network could learn the "construction rules" of a crystal family and inverse-design materials with user-specified properties. However, extending such methods to MOFs is far more challenging: MOFs contain multiple components (metal oxyhydroxy clusters and organic linkers), a wider variety of chemistry than aluminosilicate composition of zeolites, and typically larger unit cells. In fact, experimentally synthesized MOFs already contain building blocks derived from over 60 elements, and computationally simulated MOFs can include over 100 atom types, making naive atom-by-atom generation schemes computationally intractable (**Figure 1c**).[29,41,42] Generative models for MOFs therefore required leveraging their modular nature that serves as the core principle of reticular chemistry to simplify the representation and learn insights from both the computational and experimental workflows (**Figure 1d, 1e**).

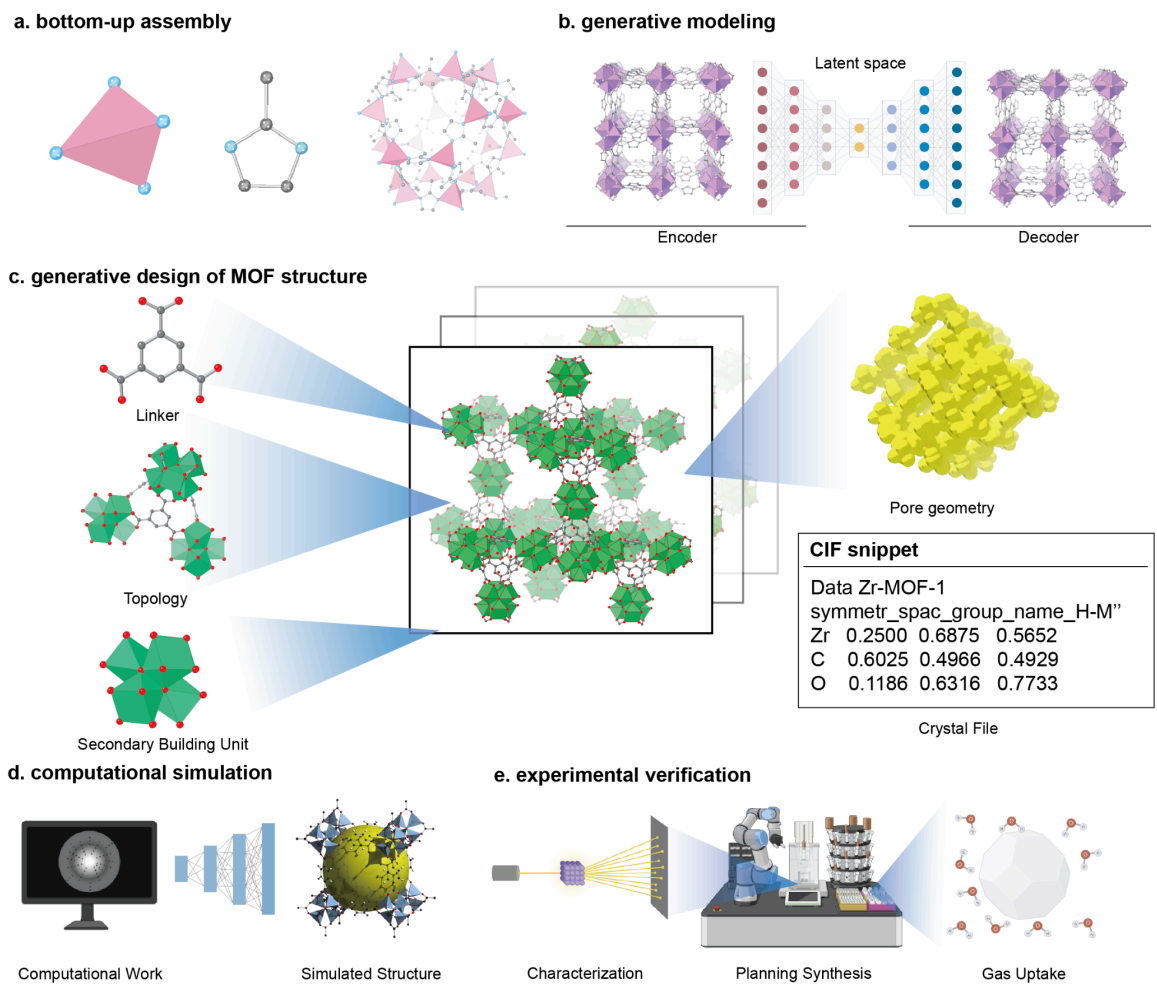

**Figure 1 | Overview of design and characterization principles of MOF.** a) Bottom-up assembly approach starting with the choices of SBUs, linkers, and stoichiometry. b) Generative approach by encoding the MOF structures in a latent space then decoding a latent vector to a MOF structure. c) Generative design of MOF structure files by selectively generating SBUs, linkers, and topologies, with optional conditions on, for example, pore geometry. d) Generated scripts from LLMs to use computational tools for simulating structures and properties of newly constructed new MOFs. e) Generated robotic actions from LLMs for experimental synthesis and measurement for designed new MOFs.

## 3. From 2D Linkers to 3D frameworks design

Early MOF discovery relied heavily on fragment-based 1D or 2D chemical representation databases and human intuition due to the reticular nature of MOFs (**Figure 2**). MOF linkers, when connectivity was considered without 3D geometry, were often specified by IUPAC names or SMILES[43] strings, and researchers searched large databases (e.g. PubChem[44,45]) for compatible organic molecules to serve as linkers for known nodes with different coordination numbers. Thus, conventional approaches for SMILES generation that have been developed for organic chemistry can be directly applied, enabling systematic *in silico* screening of MOF linker fragments and the MOF they can build up for targeted properties, for example, proposing analogues of known structures by utilizing new linkers.[46,47] Translating a 2D blueprint into a 3D crystal structure, however, is a non-trivial step in MOF chemical space due to the complexity of topological nets that connect SBUs and linkers and the periodic nature of crystalline materials. Many MOF structures significantly change upon 3D optimization, with some frameworks collapsing during the solvent removal or some breathing MOFs adjusting their pores volumes.[48–50] On the other hand, these 2D-based workflows often rely on known topologies. Recent advances in generative AI now enable direct design of 3D structures, yielding fully formed MOF crystal structures complete with realistic geometries and spatial constraints without requiring manual assembly or downstream optimization.[51] By learning from known structures and conditioning on desired properties, GenAI models have potentials of accelerating the discovery of novel frameworks with high validity rates and tailor pore environments to specific applications. Together, these models offer unprecedented flexibility for generating 3D MOF structures, reducing reliance on hand-crafted databases and enabling property-driven exploration of chemical space.

From the different components of MOFs that can be altered (i.e. topological net, inorganic node, and organic linker), the linker offers the chemical design space (**Figure 3a**). In contrast, the nuclearity and connectivity of inorganic nodes are largely dictated by reaction conditions and are therefore difficult to control directly. Leveraging this asymmetry, Park *et al.* fine-tuned the DiffLinker (a diffusion model for 3D organic molecule generation) to design organic linkers predicted to boost $CO_2$ uptake, then inserted them into a prototypical MOF architecture by fixing the metal nodes (e.g., Cu paddlewheels and Zn tetramers) and constraining the topology to the *pcu* net, which is a primitive 3-D cubic network frequently seen in MOF crystal structures.[52] This modular strategy, named GHP-MOFassemble, where GHP stands for generative high-performance, targeted carbon-capture materials without requiring generation of the entire framework at once. A set of 540 molecular fragments that frequently appeared in high-performing MOFs for $CO_2$ capture was used to allow the model to assemble over 12,000 new linker molecules, which were then used to assemble about 120,000 hypothetical MOFs. While these MOFs have yet to be synthesized and tested experimentally, the study provided clear proof-of-concept that generative AI can rapidly explore the MOF search space and pinpoint designs that would have been unlikely to emerge from human imagination or brute-force enumeration.

Beyond linker-centered design, 3D structures of MOFs can also be directly generated (**Table 1**). Yao *et al.* introduced SmVAE (supramolecular variational autoencoder), which can encode MOFs into latent vectors (**Figure 3b**), through deconstructing them into metal nodes, multiconnected organic nodes, ditopic linkers, and topological nets, and decode them back to full frameworks.[53] About 61.5% of random latent samples are chemically valid, as verified by building-block compatibility and steric-hindrance checks, meaning a majority of the materials generated by the SmVAE were chemically plausible. In addition, an auxiliary predictor was trained to steer generation in the latent space toward generating MOFs with high $CO_2$-uptake.

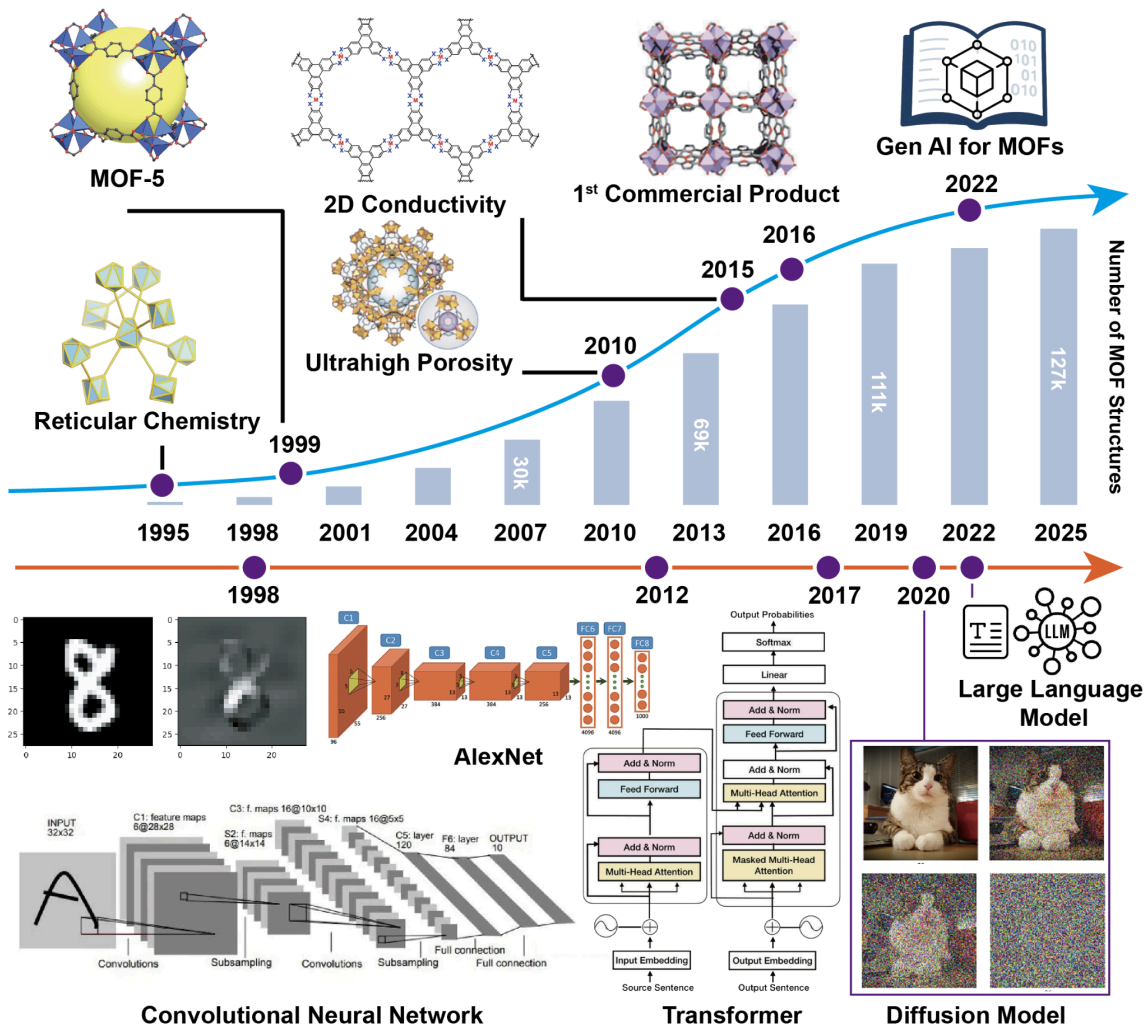

**Figure 2 | Parallel evolution of metal–organic frameworks (top) and artificial intelligence (bottom) from 1995 to 2025.** Stacked bars (right axis) show the cumulative number of MOF crystal structures (1D, 2D, and 3D) deposited in the Cambridge Structural Database (CSD) as of July 2025, where values are given in thousands (K). The upper timeline marks representative innovations in experimental MOF research,[4,54–57] while the lower timeline traces key breakthroughs in machine learning that have inspired new computational approaches to scientific discovery.[23,24,58–61]

With the popularization of diffusion models, Fu *et al.* developed MOFDiff, a coarse-grained diffusion model that constructs MOFs *de novo* by sequentially placing SBU and linker fragments together as rigid 3D objects.[62] A post-processing step aligned these fragments into a 3D periodic crystal, avoiding topological biases and producing approximately 30% novel, chemically valid frameworks. More recently, a model called MOFFUSION introduced a novel way to encode a MOF for generative tasks by using signed distance functions to represent the pore structure of the framework.[63] By training a latent diffusion model on signed distance representations of thousands of MOFs, MOFFUSION learned to generate new signed distances that correspond to plausible MOF structures. These signed distance functions were then decoded into actual MOF atomic structures via a reconstruction step, resulting in high validity of its generated structures. In addition, classifier-free guidance enables conditional generation on numeric (e.g., surface area), categorical (metal, topological net), or textual descriptors, yielding property-conditioned MOF designs.

With joint diffusion that encoded the distribution of multiple building blocks, Duan *et al.* developed building-block-aware (BBA) MOF diffusion,[64] employing an object-aware SE(3)-equivariant diffusion model that learns 3D, all-atom representations of individual building blocks while explicitly encoding crystallographic nets. BBA MOF Diffusion overcomed the size constraints of periodic materials generation and readily sampled MOFs of 1000 atoms with high geometric validity, novelty, and diversity (**Figure 3b**). A top-ranked [Zn(1,4-TDC)(EtOH)$_2$], where TDC$^{2-}$ = thiophenedicarboxylate, was synthesized and confirmed by powder X-ray diffraction (PXRD), thermogravimetric analysis (TGA), and N$_2$ sorption, demonstrating the practical utility of BBA-MOF diffusion for the design of high-performance MOFs. Building upon these advances, MOFFLOW-2 introduced a flexible two-stage framework that combines autoregressive SMILES-based generation of novel building blocks with a flow-matching model that predicts translations, rotations, and torsions for 3D assembly.[65,66] This enabled the design of chemically diverse MOFs with flexible linkers and accurate structures, overcoming the limitations of fixed libraries and rigid conformations.

**Table 1. Comparison of Key Features and Distinctions of the Major Generative MOF Models.**

| Model | Architecture | Representation | I/O | Target Properties |
|---|---|---|---|---|
| **SmVAE** | VAE (2-module encoder-decoder) | Topology + building blocks (discrete) | Input: topology + node/linker IDs; Output: same format. | $CO_2$ capture (capacity/selectivity) |
| **GHP-MOF assemble** | Diffusion (for linker gen) + rule-based assembly | SMILES linkers; fixed metal nodes & *pcu* net | Input: fragments from high-performing MOFs; Output: novel linker SMILES → assembled MOF. | $CO_2$ uptake (low-pressure); HPC screening for max capacity |
| **MOFDiff** | SE(3)-Equivariant Diffusion | Coarse-grained 3D graph (nodes = building blocks) | Input: noise; Output: block types + 3D coords. | $CO_2$ working capacity |
| **MOFFUSION** | Latent diffusion (VQ-VAE + 3D U-Net) | 3D volumetric grid | Input: Signed distance function grid; Output: new MOF signed distance function → atom structure. Conditioning: numeric, categorical, or text prompts | $H_2$ storage |
| **BBA-Diffusion** | Object-aware SE(3)-Equivariant diffusion | All-atom building blocks + net connectivity | Input: noise + topology; Output: full MOF structure in that net. Topology conditioning built-in | High-scoring candidates found via heuristics |
| **MOFFlow2** | Autoregressive transformer + Flow-matching | 3D building blocks | Input: SMILES of building blocks; Output: transition, rotation, torsion, lattice | Unconditional generation |
| **ChatMOF** | Agentic LLM + genetic algorithm | Text ↔ MOF (via internal tools) | Input: natural language; Output: predicted MOF (formula/CIF). Conditioning via prompt | $H_2$ Uptake |
| **MOFTransformer** | Multi-modal transformer encoder | Atom-based graph + methane-probe energy-grid patches | Input: MOF structure; Output: predicted properties via fine-tuned model | Gas uptake, diffusion, band gap, stability metrics |

## 4. From Native Transformers to Large Language Models

In addition to VAEs, GANs, and diffusion models, first introduced in 2017, a different type of architecture called transformers has emerged as a powerful tool for learning patterns in structured domains by using self-attention to capture contextual relationships in sequence data in the field of natural language processing (NLP).[52] Chemistry itself can be viewed as a language since atoms and building blocks follow a grammar making it well-suited to transformer-based modeling.[67] Soon after the development of landmark models like BERT[68] and generative pre-trained transformers (GPT)[63] which attained unprecedented performance in language generation tasks, the successes quickly inspired adaptations in chemistry and materials science, where sequential or graph-based representations of molecules could similarly benefit from attention mechanisms.[24,69] Transformer-based networks have been applied to molecular property prediction and reaction modeling, often outperforming traditional descriptors.[70] Naturally, transformer architectures have also been leveraged for structure–property predictions in porous crystals. A MOFormer model that encodes MOF components as text strings (e.g. "MOFid"[10] identifiers) achieved rapid property predictions without requiring 3D structures by coupling with Crystal Graph Convolutional Neural Networks (CGCNN).[71] Likewise, a multi-modal MOFTransformer pre-trained on over a million hypothetical

MOFs can be fine-tuned to predict diverse properties with improved accuracy over previous machine learning (ML) models,[72] such as gas adsorption isotherms, diffusion coefficients, and band gaps. These models not only accelerated high-throughput screening of MOFs but can also provided some interpretability (e.g. by analyzing attention weights to identify important structural features), though understanding of learned representations remains an active area of research, indicating the promise of transformers as generalizable, data-driven predictors for MOF chemistry.[73,74]

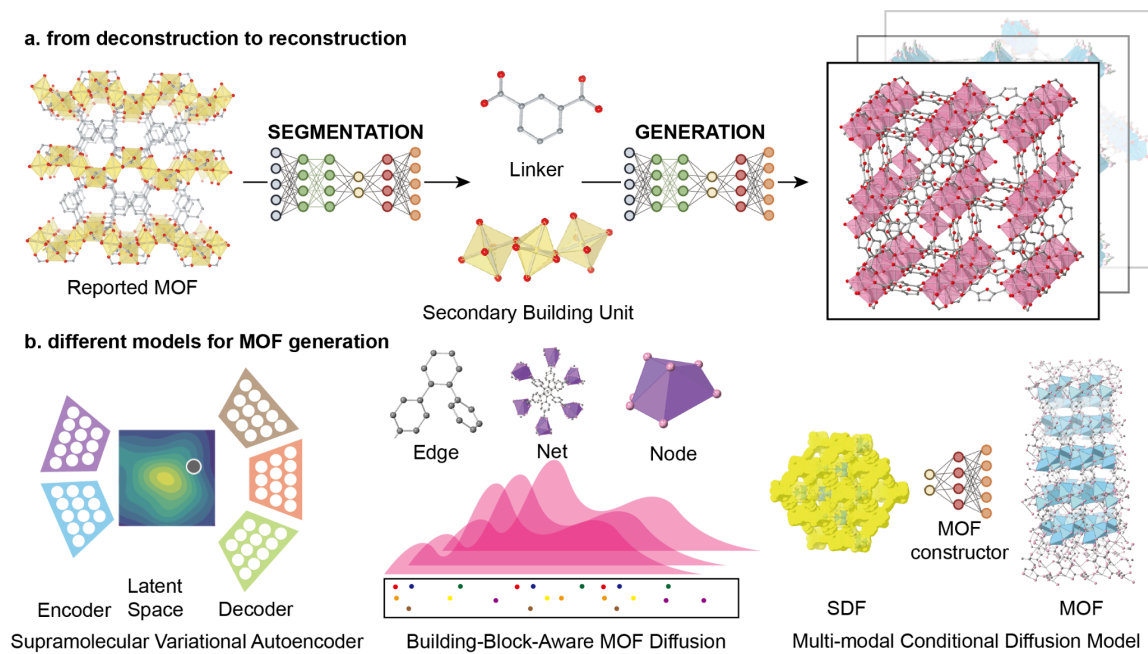

**Figure 3 | Generative design strategies for metal–organic frameworks.** a) Deconstruction-to-reconstruction workflow where reported MOF structures MOF are segmented into its organic linker and inorganic secondary building unit fragments and a generative module subsequently modifies and recombines these fragments to propose new framework topologies and atomic structures. b) Representative deep-learning architectures used for MOF generation. From left to right: a supramolecular variational autoencoder that encodes complete frameworks into a continuous latent space and decodes them back; a building-block-aware diffusion model that samples edges, nets and nodes to assemble MOFs stochastically; and a multi-modal conditional diffusion model that generates signed-distance-function (SDF) volumes which are subsequently converted into full atomic MOF structures.

Beyond prediction tasks, general-purpose large language models (LLMs) have demonstrated remarkable generative capabilities that are now being repurposed for chemical and materials innovation. In their native domain of NLP, transformer-based LLMs can compose fluent passages of text or even generate computer code, illustrating their capacity for open-ended synthesis of information. This generative power has been harnessed in chemistry for tasks like *de novo* molecular design, reaction pathway generation, synthesizability prediction, and even crystal structure proposal.[75–78] For instance, language models fine-tuned on chemical corpora can auto-complete SMILES strings to propose novel compounds or suggest multi-step synthetic routes, effectively treating molecular design as a language-generation task.[79] Similarly, language models fine-tuned on SMILES string and IUPAC names of existing reported MOF linkers demonstrate the ability to propose new and valid organic linkers based on human instruction.[80] Furthermore, coupled with evolutionary algorithms, the internal chemistry and materials knowledge and reasoning capability of LLMs were exploited for functional organic molecules,[76] transition metal complexes,[81] and bulk materials design,[77] performing on par or better than conventional genetic algorithms and Bayesian optimization.[82] More recently, a text-guided GenAI model, Chemeleon, conditions a diffusion-based crystal generator on textual descriptions, allowing it to propose novel crystal structures informed by both chemistry language and 3D structural data.[77] Together, these research findings suggest that LLMs can serve not only as natural language interfaces, image creators, or coding assistance popularized for commercial use, but also as creative engines in MOF research for generating new MOF building block combinations, suggesting synthesis plans, and integrating domain knowledge on-the-fly to accelerate the discovery of next-generation MOFs.

**5. Human-in-the-Loop Workflows for Gen AI-Driven MOF Discovery**

Generative modeling provides an unprecedented way of proposing new candidate materials, potentially unlocking chemical spaces unimaginable by enumerative library screening. Generative models alone, however, are only the beginning rather than the end of a materials discovery pipeline (**Figure 4**). Once obtaining new MOF structures, we need to answer crucial questions: *Are these structures physically plausible and synthesizable? Do they have the properties we want? How do we actually make them in the lab?* Therefore, building end-to-end MOF discovery workflows that couple generative models with simulation, optimization, and robotic experimentation are desirable for reticular materials generation with targeted properties.[2,25,83,84]

Traditional MOF discovery relies on manual workflow, where researchers search and interpret literature, plan experiments, and conduct synthesis and characterization.[2,25,26] These approaches, while effective, are limited by human cognitive biases and throughput constraints, making them inadequate for exploring the vast combinatorial MOF chemical space.[85] In contrast, interactive GenAI platforms now let chemists exploit data-driven literature mining without bespoke coding.[86,87] Some early examples of interactive GenAI for MOF research include the ChatGPT Chemistry Assistant[88], Paragraph2MOFInfo[89], LLM-NERRE[90], Eunomia,[91] and L2M3[92], which helps extract MOF data from the literature, allowing experimentalists to query papers and synthesis procedures conversationally via natural language. By unifying literature-derived insights with computational chemistry data and linking them to crystal structures, LLM-based workflows such as MOF-ChemUnity[93] enable literature-informed AI assistants for scalable, multisource data mining. In the meantime, multimodal generative models with vision capabilities, such as GPT-4V have also shown promise in helping labeling and extracting MOF characterization data from literature figures.[94] On the other hand, agentic-AI is a promising approach to augment LLMs with proper tools for task execution, such as data mining[91] and generation. For example, ChatMOF[95] developed by Kang and Kim demonstrated a tool-augmented system where an LLM agent marshals tools for database search, property prediction, and structure generation, responding to plain-language requests, such as *"Can you suggest a MOF structure with a pore size around 1.2 nm and high methane storage capacity?"* The agent interfaces with provided generative modules, ranging from genetic algorithms to diffusion models, to propose novel MOF structures.

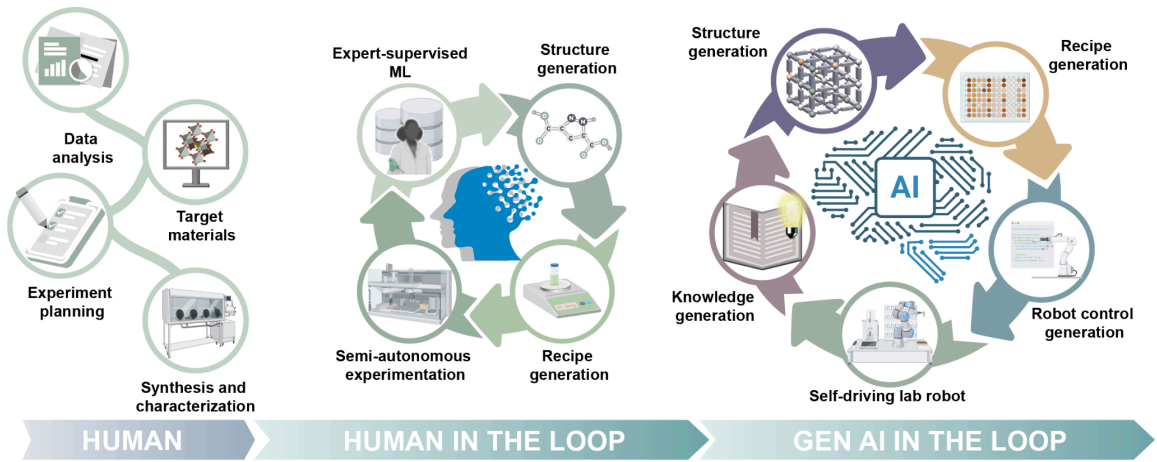

**Figure 4 | Evolution of design and synthesis practices for MOF innovations.** Schematic illustration of the progressive integration of generative AI into close-looped MOF discovery. The transition from human-led decision-making (left and middle) to AI-driven orchestration is enabled through structure, recipe, instruction, and knowledge generation modules (right).

Beyond data mining and inferencing, GenAI agents have the potential to integrate with enhanced-sampling simulations to form a self-optimizing, closed-loop pipeline for MOF discovery. An autonomous agent could coordinate multi-fidelity strategies (e.g., nested sampling[96], metadynamics[97], thermodynamic maps[98]) to act as a physics-informed oracle. Inexpensive coarse evaluations could outline the free-energy landscape, while high-resolution profiles would refine priority frameworks to capture features like sorption-induced phase transitions[99] and gate-opening events[100] that influence usable capacity. This approach mirrors multifidelity Bayesian optimization[101], filtering candidates efficiently and reserving costly evaluations for top performers. Besides, it is envisioned that thermodynamic observables ($\Delta H_{ads}$, $\Delta S_{ads}$) and phase boundaries be introduced to suggest optimal linker-node-topology combinations that aim to maximize gas uptake, selectivity, and structural robustness under realistic operating conditions. Leading candidates could be synthesized by automated platforms and validated through high-throughput characterization, feeding data back into the workflow.

Closer to the experimental end, Yan *et al.* developed MOFA[102], a workflow that couples a generative linker model (MOFLinker) with automated assembly, simulation-based screening, and on-line retraining, forming a self-improving loop for MOF discovery. They found MOF candidates whose predicted $CO_2$ uptakes (GCMC at 0.1 bar, 300 K) ranked in the top decile of a 137,652 hypothetical MOF benchmark. Despite these advances, experimental verification remains predominantly manual, with researchers synthesizing and characterizing a limited subset of AI-generated candidates constrained by synthetic feasibility and available resources.[103–105] Moreover, even successfully synthesized MOFs often require post-synthetic functionalization to achieve optimal performance. Those modifications are difficult to model in silico and represent an additional layer of complexity in translating AI-generated designs to functional materials. To address these limitations, echoing developments in LLM-driven organic synthesis (e.g., ChemCrow[106] and Co-Scientist[107]), MOF research is beginning to incorporate LLM-copilot workflows such as the ChatGPT Research Group[108] and Chemical Robotic Explorer[109] for iterative reaction optimization. These systems reveal that the application of GenAI models extends beyond merely structure generation as they can also generate scripts for synthetic protocols and robotic execution. Recently, Inizan *et al.*[51] developed an agentic AI system to propose and validate synthesizable MOFs. Here, an LLM is first used to propose novel MOF compositions, followed by a diffusion model to generate crystal structures, and quantum chemistry calculations to optimize and filter candidates. Five AI-predicted MOFs were experimentally synthesized with synthesis conditions systematically explored using a robotic high-throughput platform.

As laboratory automation continues to evolve, it is envisioned that GenAI-in-the-loop workflows will be increasingly viable. In such systems, AI agents autonomously generate candidate structures, design synthesis protocols, interface with robotic platforms, and iteratively refine outputs based on experimental feedback. This paradigm represents a shift from static and manual workflows towards adaptive and closed-loop pipelines that can significantly accelerate MOF discovery.

**6. Outlook**

Generative AI is poised to fundamentally reshape how we discover MOFs, transforming the process from a largely manual, intuition-driven craft into a data-rich, AI-accelerated science. To draw an analogy, reticular chemistry gave us the chemical language and grammar to construct MOFs deliberately, rather than by serendipity. Now, generative AI is like a creative agent fluent in that language, brainstorming new sentences (MOF structures) that are grammatically correct (chemically valid) but novel and inventive. The collaboration between humans and machines can amplify innovation: humans set the objectives and constraints inspired by application needs, and AI explores the possibilities at a speed and breadth that is close to what an entire team of human experts can achieve in years.

To fully realize the potential of GenAI, several challenges still remain ahead (**Figure 5**). First, validation bottlenecks will arise as AI generates more candidates than we can reasonably test experimentally. Manually synthesizing and evaluating a predicted MOF can take months, and might result in a failure due to inaccurate scoring functions. This is where improved screening models and autonomous experimentation must step in. Second, experimental constraints pose significant integration challenges. MOF discovery often requires new experimental protocols or customized hardware, limiting the applicability of existing robotic platforms. Developing flexible and modular self-driving laboratories capable of accommodating diverse synthesis routes is therefore critical. In parallel, GenAI must also evolve to generate robotic workflows tailored to the constraints of individual systems. Third, diversity vs. realism will potentially be a perpetual trade-off: pushing a model too hard to explore out-of-distribution MOFs may result in the model outputting unphysical structures, but constraining it too tightly will lead to recapitulation of known MOFs. This also raises the issue of hallucination, where generative models may confidently create structures or synthetic pathways that are chemically implausible. Addressing these issues requires integrating domain knowledge and human-in-the-loop strategies to penalize unrealistic generations. Future work on techniques like reinforcement learning with human chemist feedback could allow MOF researchers to iteratively refine generative models by giving feedback on small subsets of outputs. Community standards, including unified benchmarking metrics that reflect practical real-world benefits, and sharing of models will also be important. Indeed, many groups have released their code and even pretrained models.[62,63,72,95] Democratizing the toolsets and making them easily accessible (e.g. through language model interfaces) will enable researchers, especially synthetic chemists, who are not AI specialists, to apply generative design to their specific MOF problems.

While most generative models leverage repeatable building blocks to enable scaling to hundreds of atoms, limiting the available substructures to known motifs in the training data hinders the discovery of truly novel structures, including previously unknown linkers, metal nodes, and topologies. Models operating directly on atoms, such as All-Atom Diffusion Transformers[62], are promising for small molecules and crystals, but lag behind building block-based approaches in terms of validity for MOF generation. An important future direction of research will be combining the flexibility of all-atom methods with the coarse-grained nature of building block-based models, which could furthermore incorporate transfer learning to leverage knowledge about crystals and molecules.

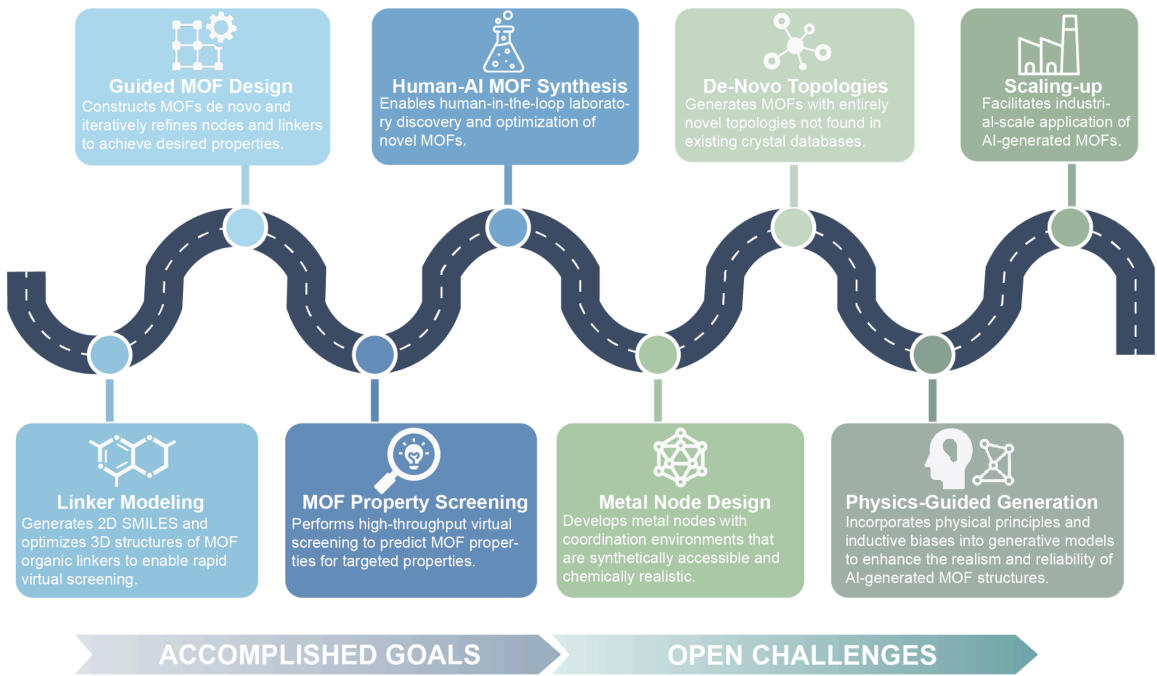

**Figure 5 | Accomplished goals (blue) and open challenges (green) towards realization of generative-AI designed MOFs.**

It is also worthwhile noting that another critical limitation lies in the quality of available structural data. While hypothetical MOF datasets such as hMOF[9,110], ToBaCCo[38,111,112], and other *in silico* libraries provide valuable inputs for training generative models, they often lack experimental validation. By contrast, databases like CoRE MOF[113–115] and QMOF[111,112] are rooted in experimental data but may not cover MOFs that are harder to crystalize and can suffer from potential bias of structural false negatives. With this data limitation, how generative models can be designed to generate novel metal nodes and topology is an open challenge. In addition to previously raised concerns about geometric and charge errors,[116] recent work found over 40% of widely used "computation-ready" MOFs contain flawed metal oxidation states,[117] undermining their reliability for DFT-level modeling and generative workflows. These findings underscore the importance of structure validation and repair pipelines that are not just for simulations, but also for ensuring generative models are trained on physically meaningful data. Moreover, MOF chemical data is inherently multimodal and derived from diverse sources. While current efforts in generative AI focus primarily on crystal structure generation, incorporating additional data modalities such as X-ray diffraction patterns could strengthen links between computational modeling and synthesis.[118] Realizing this vision will require holistic datasets that integrate computational results, experimental measurements, and human-curated insights.[93]

One exciting prospect is the integration of physics-based priors into generative models. Imagine a generative model that inherently respects force-field energetics or DFT stabilities, which could be approached by combining energy-based modeling with deep learning, so that the model preferentially generates structures in low-energy configurations.[119,120] Some initial work using score-based generative models with physical constraints is appearing in molecular domains;[121,122] for crystals, the idea of crystal symmetry and energy could be baked into the model's training objectives. Community datasets that combine atomic coordinates[112] with comprehensive energetics, including enthalpic and entropic contributions, could support transfer-learning workflows for adsorption[74] and catalysis[123], thereby facilitating each discovery cycle and enriching the collective knowledge base.

Furthermore, this knowledge can also be incorporated post-training. Reinforcement learning-based fine-tuning with physics-based reward models and inference-time control using reward guidance are two promising directions. As available data will be run out for training generative models, inference-time scaling can be key to further improving model performance. As generative models produce candidates that push into unknown chemistry regimes, close collaboration with experimentalists will be essential to identify which suggestions are truly synthesizable. Furthermore, experimentalists will be able to verify the properties of the proposed new materials once obtained. Looking forward, this paradigm shift invites new thinking on how success in GenAI-guided MOF discovery should be defined, whether through the accurate synthesis of predicted structures, the generation of experimentally testable hypotheses, or the continual refinement of foundation models with new knowledge. Establishing robust success metrics will be essential for benchmarking progress and guiding future trajectory of MOF discovery in the GenAI era.

In reflecting on the journey from early initial trial-and-error MOF designs to today's AI-driven discovery, one is reminded of how data-driven insight can complement fundamental knowledge. Reticular chemistry gave us principles like "the topology can be decoupled from the components" and "infinite nets are achievable via strong bonds."[2,124] Generative AI doesn't change those principles; in fact, it leverages them.[25,26] Reticular materials provided an ideal playground for GenAI, as it is a domain with modular chemistry and a wealth of data, and AI, in turn, is giving back to MOF science by revealing how vast and varied that playground really is. As the field progresses, we anticipate seeing AI-designed MOFs setting new records: perhaps an ultraporous MOF that captures and catalytically breaks down methane from agricultural or industrial sources, a water harvesting MOF with high capacity and fast kinetics providing clean water solution for semi-desert areas, or a $CO_2$-capturing MOF that works efficiently in humid air and makes direct air capture (DAC) economically viable. In fact, in recent years companies like BASF and Svante are already advancing newly discovered, bench-developed MOFs toward industrial production.[8] In conclusion, the MOF of the future might very well be conceived by an AI, but it will be the human scientists who guide the AI, synthesize the creation, and ultimately implement it in solving real-world problems. In that sense, generative AI is not replacing the reticular chemist. In fact, it is empowering them, much like a powerful new synthesis platform or characterization tool would. The reticular revolution has entered the digital age, and the prospects for innovation have never been more exciting.

**Competing interests**

The authors declare no competing interests.

**Acknowledgements**

Z.Z. acknowledge the support provided by the EQT Foundation Breakthrough Science Grants. A.N. gratefully acknowledges support from the Eric and Wendy Schmidt AI in Science Postdoctoral Fellowship, a Schmidt Sciences, LLC program. Z.Y. and S.S. acknowledge support by the National Science Foundation through University of Washington Materials Research Science and Engineering Center, DMR-2308979. A.A.-G. thanks Anders G. Frøseth for his generous support. A.A.-G. and Y.K. acknowledge the generous support of Natural Resources Canada and the Canada 150 Research Chairs program. A.A.-G. and V.B. are supported by the University of Toronto's Acceleration Consortium, which receives funding from the CFREF-2022-00042 Canada First Research Excellence Fund. Y.K. was supported by the CIFAR AI Safety Catalyst Award (Catalyst Fund Project #CF26-AI-001).